\documentclass[11pt,twoside]{article}

\usepackage{asp2004}
\usepackage{epsf}
\usepackage{psfig}
\usepackage{lscape}

\begin{document}

\title{Large Scale Structure at 24 Microns in the SWIRE Survey}

\vspace{-5mm}

\author{Frank J. Masci and the SWIRE Team}

\affil{{\it Spitzer} Science Center, California Institute of Technology,
220-6, Pasadena, CA, 91125}

\begin{abstract}
We present initial results of galaxy clustering at 24$\micron$
by analyzing statistics of the projected galaxy distribution
from {\it counts-in-cells}. This study focuses on the ELAIS-North1
SWIRE field. The sample covers $\simeq5.9\,{\rm deg}^2$ and
contains 24,715 sources detected at $24\micron$ to a $5.6\sigma$
limit of 250$\mu$Jy (in the lowest coverage regions).
We have explored clustering as a function of
3.6 - 24$\micron$ color and 24$\micron$ flux density
using angular-averaged two-point correlation
functions derived from the variance of counts-in-cells on scales
$0\deg.05-0\deg.7$. Using a power-law parameterization,
$w_{2}(\theta)=A(\theta/{\rm deg})^{1-\gamma}$, we find
[$A,\,\gamma$] = [$(5.43\pm0.20)\times10^{-4},\,2.01\pm0.02$]
for the full sample ($1\sigma$ errors throughout). We have inverted Limber's
equation and estimated a spatial correlation length of
$r_{0}=3.32\pm0.19\,h^{-1}$Mpc for the full sample, assuming
stable clustering and a redshift model consistent with
observed $24\micron$ counts.
We also find that blue [$f_\nu(24)/f_\nu(3.6)\leq5.5$] and red
[$f_\nu(24)/f_\nu(3.6)\geq6.5$] galaxies have the lowest and highest
$r_{0}$ values respectively, implying that redder galaxies
are more clustered (by a factor of $\approx3$ on
scales $\ga0\deg.2$). Overall, the clustering estimates
are smaller than those derived from optical surveys,
but in agreement with results from IRAS
and ISO in the mid-infrared. This extends the notion to
higher redshifts that infrared selected surveys show weaker
clustering than optical surveys.
\end{abstract}


\vspace{-11mm}

\section*{Analysis and Summary}

\vspace{-3mm}

The {\it Spitzer} Wide-area Infrared Extragalactic legacy program
\citep[SWIRE;][]{Lonsdale.etal03} is expected to detect over two
million galaxies at infrared wavelengths from 3.6 to $160\micron$
over $49\,{\rm deg}^{2}$ and to redshifts $z\simeq2.0$.
Our ELAIS-N1 sample corresponds to a completeness of $\simeq90$\%
and we used a $\simeq5.9\,{\rm deg}^2$ central
region where the completeness was uniform to within $\pm0.5\%$.
Reliability of each $24\micron$ detection was ensured by
requiring an IRAC $3.6\micron$ association with SNR $\geq\,10\sigma$
(where $0.6\leq\sigma/\mu{\rm Jy}\leq1.0$) and separation
$\leq 2\arcsec$. Stars were identified and rejected using a combination
of the $3.6\micron$ stellarity index ($\geq0.9$) and cuts in
$f_\nu(3.6)$ versus $f_\nu(24)/f_\nu(3.6)$ flux ratio
\citep[see][]{Masci.etal05}.
This study is the first of its kind at this wavelength and sensitivity,
reaching a factor of $\approx1000$ deeper in flux density than the IRAS
$25\micron$ galaxy surveys.

The method of counts-in-cells
\citep[CICs; see][and references therein]{Masci.etal05}
provides the full galaxy count distribution function
within a cell of given size, and its
moments are related to the classical $n$-point correlations.
The variance in particular is related to the
cell-averaged {\it two-point} correlation function. The advantage of CICs
over traditional, direct binning methods is that the data does not
require binning, moments from CICs have better signal-to-noise ratio properties,
no random comparison sample is needed, and systematics
from catalog boundaries and finite sampling are better handled.

\begin{figure}
\vspace{-3mm}
\plottwo{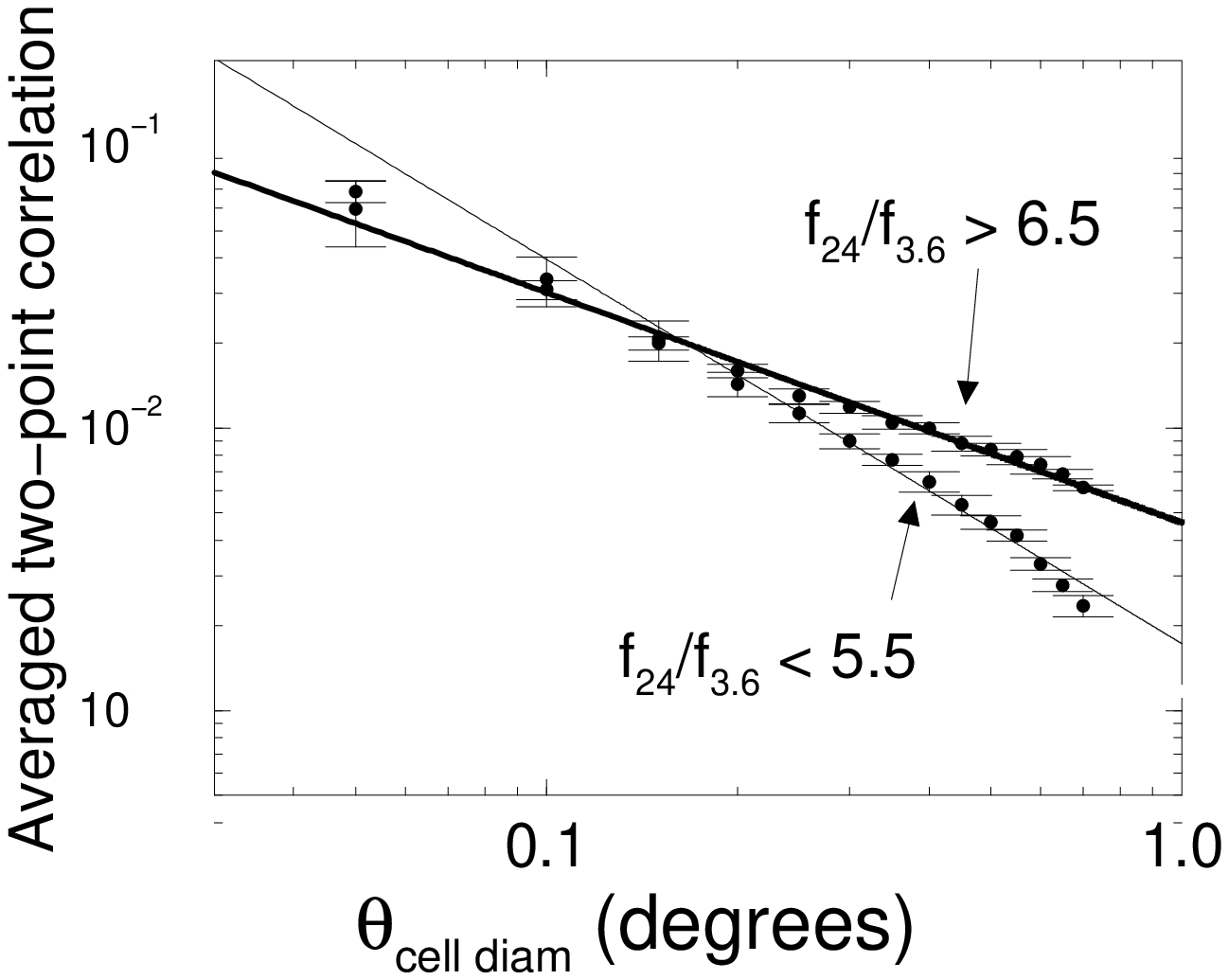}{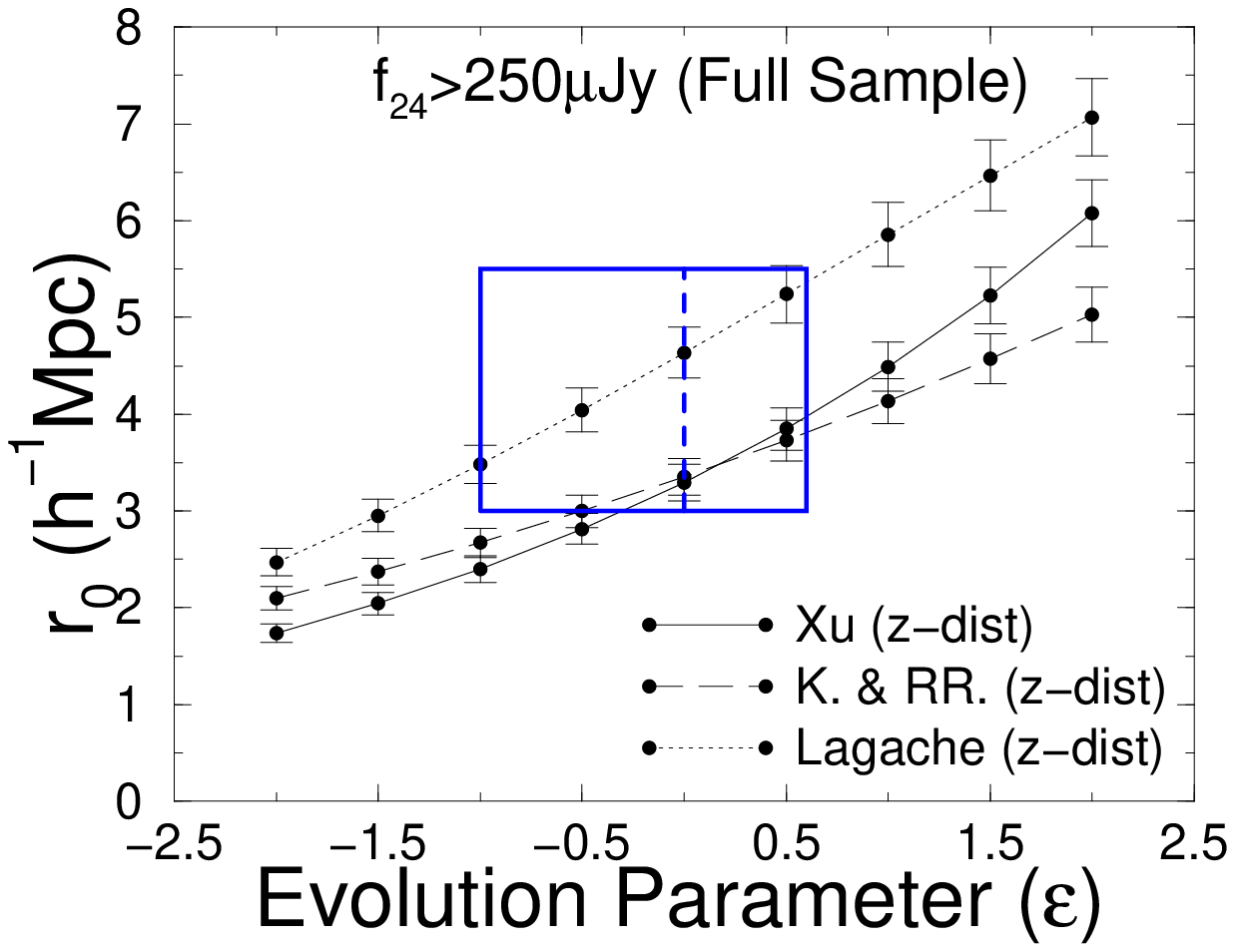}
\caption{Angular-averaged correlation versus angular scale for two
color subsamples (left) and correlation length versus the
evolution parameter $\epsilon$ (right), assuming three different model
redshift distributions. The rectangle encloses a region consistent
with predictions from N-body simulations and other observations.\label{fig}}
\end{figure}

Figure \ref{fig} (left panel) shows the angular-averaged correlation function
as a function of cell diameter. The angular range corresponds to
{\it comoving} scales of $\simeq2-24\,h^{-1}$Mpc at
the expected median redshift of $\simeq 0.8$ for the full sample,
as predicted from the \citet{Xu.etal01} model and
$(\Omega_{m},\,\Omega_{\Lambda}) = (0.3,\,0.7)$.
The striking feature
is that redder (mid-IR excess) galaxies exhibit a stronger projected
clustering than blue galaxies on scales $\ga0\deg.2$.
This difference is difficult to explain by varying projection effects
since their {\it photometric} redshift distributions
(although incomplete) are not too dissimilar. More work is needed to
verify this.

We have estimated correlation lengths
by inverting Limber's equation with
three different model redshift distributions and assuming
a simple evolution model for 3D clustering:
$\xi(r,z)=(r/r_{0})^{-\gamma}(1+z)^{-(3+\epsilon)}$,
where $r$ is a proper coordinate and $\epsilon=0$
implies constant (stable) clustering in proper coordinates.
We found spatial correlation lengths in the range
$r_{0}\simeq3.3$ to $4.7\,h^{-1}$Mpc for the full sample
assuming $\epsilon=0$ (right panel in Fig.~\ref{fig}). For the
redshift model of \citet{Xu.etal01},
whose predictions agree remarkably well with observed $24\micron$ counts
\citep[e.g.,][]{Chary.etal04}, we find $r_{0}=3.32\pm0.19\,h^{-1}$Mpc. 
We also found that red galaxies are more clustered
by a factor of $\sim2.6$ in their spatial correlation amplitude on
$5\,h^{-1}$Mpc scales at $z=0$ than blue galaxies (not shown).
This may suggest that merger driven
starbursts are dominating the mid-infrared excess population
at $z\ga0.5$. We note however that our 3D clustering estimates depend
strongly on the assumed redshift distribution. This reinforces the need
to establish the redshift distribution of SWIRE galaxies.

\acknowledgements

Support for this work was provided by NASA through an award
issued by the Jet Propulsion Laboratory, California Institute of
Technology under NASA contract 1407.

\end{document}